# On the Possibility of Developing Quasi-CW High-Power High-Pressure Laser on 4p–4s Transition of ArI with Electron Beam - Optical Pumping: Quenching of 4s ($^3P_2$) Lower Laser Level

**A A Ionin\*, I V Kholin, A Yu L'dov, L V Seleznev,
N N Ustinovskii, and D A Zayarnyi**

*P.N.Lebedev Physical Institute, Leninsky prospect 53, 119991 Moscow, Russia*
*\*e-mail*: aion@sci.lebedev.ru

**Abstract.** A new electron beam-optical procedure is proposed for quasi-cw pumping of high-pressure large-volume He–Ar laser on $4p[1/2]_1 – 4s[3/2]_2$ argon atom transition at the wavelength of 912.5 nm. It consists of creation and maintenance of a necessary density of $4s[3/2]_2$ metastable state in the gain medium by a fast electron beam and subsequent optically pumping of the upper laser level via the classical three-level scheme using a laser diode. Absorption probing is used to study collisional quenching of $Ar^*$ metastable in electron-beam-excited high-pressure He–Ar mixtures with a low content of argon. The rate constants for plasma-chemical reactions $Ar^*+He+Ar$-$Ar_2^*$+He $(3.6 +- 0.4) \times 10^{-33}$ cm$^6$/s, $Ar+2He$-$HeAr^*$+He $(4.4 +- 0.9) \times 10^{-36}$ cm$^6$/s and $Ar^*+He$-Products+He $(2.4 +- 0.3) \times 10^{-15}$ cm$^3$/s are for the first time measured.

## 1. Introduction

A high-pressure He–Ar gas laser at $4p[1/2]_1 – 4s[3/2]_2$ ArI transition with lasing wavelength of 912.5 nm was for the first time launched more than forty years ago [1]. This laser pumped by a fast high-voltage subnanosecond electric discharge is a classical representative of lasers on "self-terminating transitions" and has a very low lasing efficiency. The low efficiency is due to the fact that there is no effective mechanism for depopulating the lower laser level in such lasers. The metastable level in this He–Ar laser is depopulated by rather slow quenching in collisions with gain medium atoms. A development of lasing at such a transition leads to populating the lower laser level quite rapidly and, consequently, to lasing termination.

In papers [2–4], a high-pressure He–Ar laser with combined two-step pumping was put into practice. At the first stage, a sufficient density of $4s[3/2]_2$ metastable Ar states was produced by conduction electrons in high-current electric discharge plasma. In the second stage, a population inversion on the laser transition was formed in accordance with the classical three-level scheme due to optical excitation of $4p[5/2]_3$ higher-lying level and subsequent collisional quenching to the upper laser level $4p[1/2]_1$. Optical pumping was accomplished with a laser diode via allowed optical transition $4p[5/2]_3 - 4s[3/2]_2$ with a large oscillator strength and the wavelength of 811.5 nm that is relatively close to the laser wavelength.

In [1–4], pumping of high-pressure gain medium by an electric discharge imposed fundamental limitations on the maximum values of excited volume and pump pulse width. Indeed, their increase over tens of cubic centimeters and tens of nanoseconds, respectively, lead to the development in rare gas mixtures of various, primarily ionization, instabilities because of particular structure of rare gas atomic levels (small energy gap between the lowest excited states and ionization potential), which results in rapid discharge contraction. To bring into action a quasi-cw operation with a large gain volume, we propose to create and maintain the necessary density of excited metastable atoms in the active medium during a required time by a fast electron beam (figure 1).

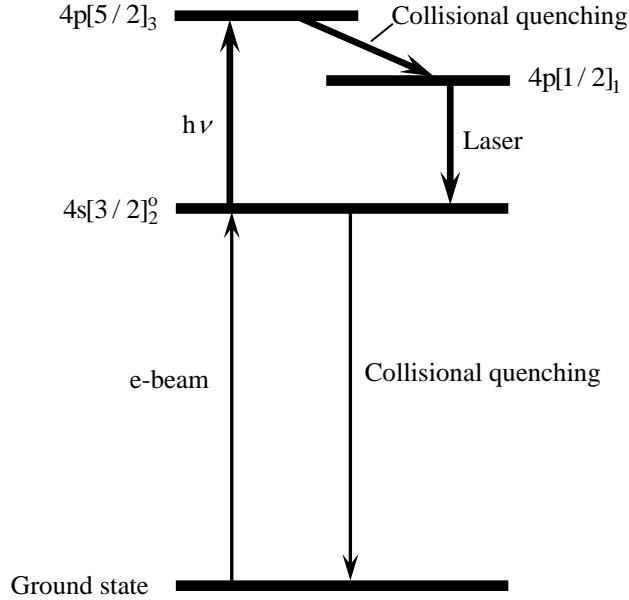

**Figure 1.** Schematic of combined e-beam (thin arrow) and optical pumping (thick arrow).

Previously we employed a similar approach to develop a high-power electro-ionization (i.e. e-beam sustained discharge) Ar–Xe laser on 5d–6p XeI transitions [5, 6]. A fast electron beam was used to ionize gas mixture producing sufficient density of metastable 6s states of xenon and, thus, forming a gain medium with new electrical properties. Applying an electric field to the ionized gain medium with such new properties made it possible to efficiently pump 5d–6p laser transitions via a four-level excitation scheme not from the ground state, but from the 6s states already produced by the electron beam, which brought Ar–Xe lasers to a qualitatively new level setting them on a par with the most powerful of the known lasers [5–8]. In the present paper, we also propose to form a gain medium with new properties by means of fast electron beam, which makes it possible to perform a quasi-cw (in contrast to [2–4]) optical pumping of ArI $4p[1/2]_1$ - $4s[3/2]_2$ transition through a three-level scheme of figure 1 from metastable $4s[3/2]_2$ state. Implication of such a quasi-cw "electron beam-optical" procedure has to highly increase both geometric dimension of the gain medium and laser pulse width and, consequently, substantially elevate the laser output.

To estimate an efficiency of the proposed scheme before carrying out a complicated experiment, it is necessary to know, in particular, rate constants for the collisional plasma-chemical reactions in He–Ar mixtures. However, relevant literature data are currently practically absent. This study is the first step in this direction and deals with collisional quenching of the $4s[3/2]_2$ lower laser level in following reactions:

$$\mathrm{Ar}(4s[3/2]_2^o) + \mathrm{He} + \mathrm{Ar} \rightarrow \mathrm{Ar}_2^* + \mathrm{He}, \tag{1}$$

$$\mathrm{Ar}(4s[3/2]_2^o) + 2\mathrm{He} \rightarrow \mathrm{HeAr}^* + \mathrm{He}, \tag{2}$$

$$\mathrm{Ar}(4s[3/2]_2^o) + \mathrm{He} \rightarrow \mathrm{Products} + \mathrm{He}. \tag{3}$$

In our research He–Ar mixtures excited by a beam of fast electrons were experimentally studied. Measurements of the reaction rate constants were performed by the method of absorption probing based on the dependences of the decay rate of $4s[3/2]_2$ state on the total gas pressure and concentration ratio between the working and buffer gas. For this purpose, the absorption time behavior of a probing light pulse at the wavelength of 912.3 nm (figure 2) in the afterglow of gas mixture excited by a high-power beam of fast electrons was recorded.

## 2. Experiment

Our experiments were carried out with a cold-cathode electron gun being a part of the pulsed laser facility "Tandem" [17]. An electron beam of $5 \times 100$ cm$^2$ cross section with a peak current density up to 1.5 A/cm$^2$, bell-shaped current pulse with full width of ~2.5 $\mu$s and electron energy of ~250 keV was injected into the measuring chamber with an active volume of $5 \times 5 \times 100$ cm$^3$ perpendicular to its optical axis. This 5-*l* measuring chamber was made of stainless steel. Before filling, the measuring chamber was evacuated through a liquid nitrogen trap to a residual pressure of ~$10^{-5}$ Torr. Gas leakage into the evacuated chamber and its walls outgassing does not exceed $10^{-3}$ Torr/hour. Helium of grade "A" (99.998 vol %) and argon of high-purity grade (99.9992%) in gas mixtures with concentration ratio He–Ar = 50:1, 75:1, 100:1 and 200:1 at the total gas pressure ranging from 1.75 to 4.0 atm were used in the experiments.

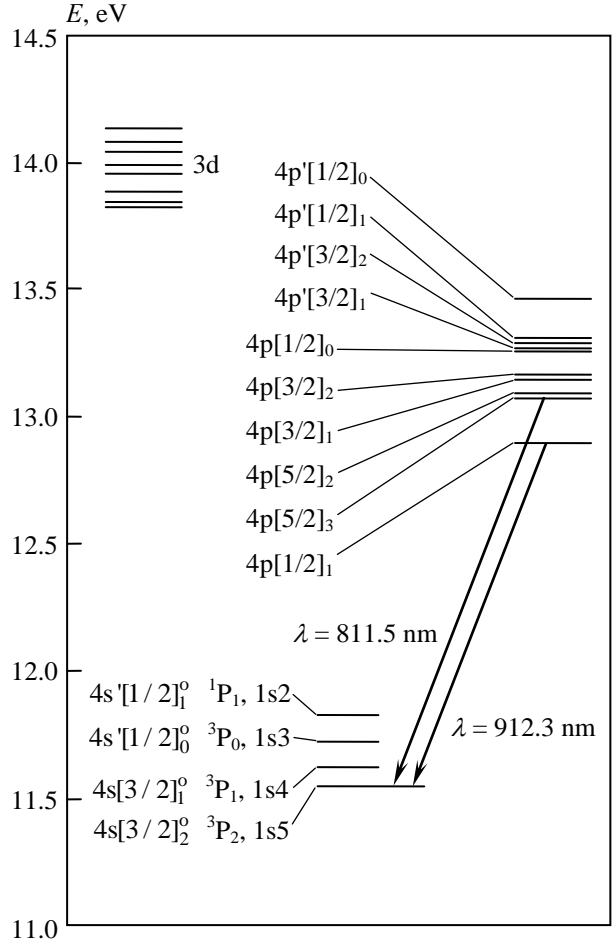

**Figure 2.** Diagram of the low-lying excited levels of Ar atom.

The optical scheme of our measurements is shown in figure 3. A broadband ISI-1 (*1*) light source with a pulse width of ~30 μs (figure 4, curve *B*) served as a source of the probing signal. The light was collimated into a nearly parallel beam of probe radiation 5 cm in diameter. After passing through the e-beam-pumped gas mixture in chamber *5* it was focused onto the entrance slit of high-aperture monochromator MDR-2 (*8*) equipped with a 600 grooves/mm diffraction grating. The radiation selected by the monochromator tuned to the needed wavelength was detected by photodetector *10* comprised of a high-speed PIN-photodiode BPW34 of *Infineon Technologies AG* and AD8055 broadband operational amplifier from *Analog Devices* housed in a metal case with double shielding. A fraction of the radiation reflected from plane-parallel glass plate *4* set in front of the measuring chamber was directed, bypassing the measuring chamber, onto other MDR-2 monochromator *9* and photodetector *11*. To suppress a short-wave radiation going on the photodetectors in the second diffraction order, an optical filter KS-10 (*3*) was set into the probing radiation beam. Signals from the photodetectors were monitored using a two-channel PC-based digital storage oscilloscope DSO-2010 of *Link Instruments Inc.*

Thus, the measurement scheme allowed us to simultaneously monitor the probing pulse shape both before and after passing the absorbing medium with a temporal resolution not worse than 100 ns. In the afterglow of gas mixture excited by the electron beam pulse, when processes of recombination and relaxation from high-lying states are completed, population of the state under the study is to be governed, basically, by quenching processes in reactions (1)–(3):

$$\frac{d[\text{Ar}^*]}{dt} = -k_1[\text{Ar}][\text{He}][\text{Ar}^*] - k_2[\text{He}]^2[\text{Ar}^*] - k_3[\text{He}][\text{Ar}^*]. \tag{4}$$

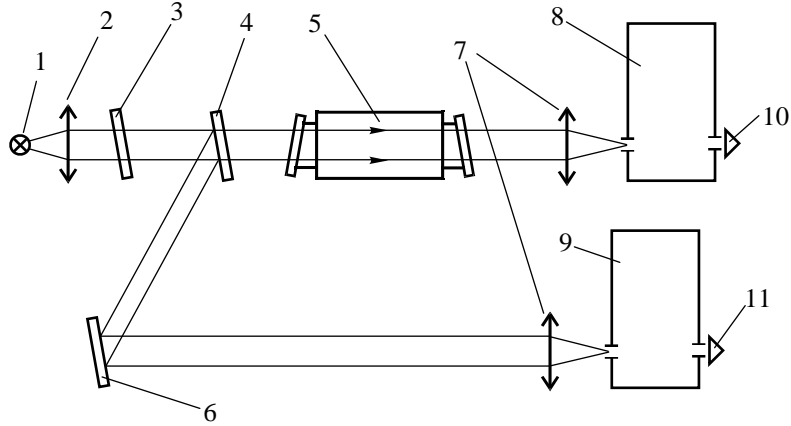

**Figure 3.** Layout of the absorption measurements: (1) ISI-1 pulsed light source; (2) collimating lens; (3) KS-10 cut-off optical filter; (4) beam splitter; (5) measuring chamber; (6) turning mirror; (7) focusing lenses; (8, 9) high-aperture MDR-2 monochromator; (10, 11) photodetectors.

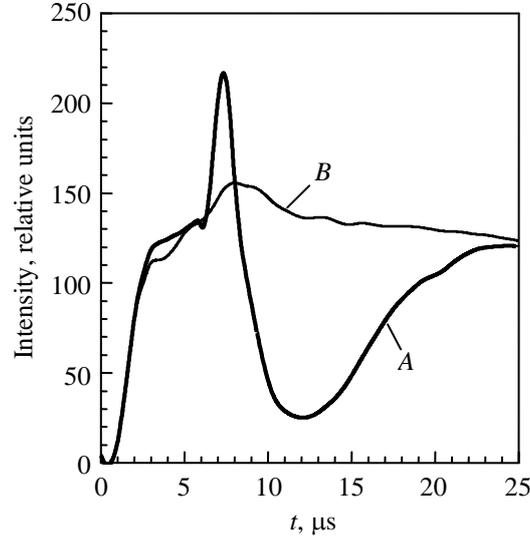

**Figure 4.** Time behavior of 912.3-nm probe pulse intensity passed through (*A*) and beyond (*B*) the excited mixture (He–Ar = 100:1) at gas pressure of 2.5 atm. The intensity spike on signal *A* curve before beginning of the absorption is due to fluorescence on the transition under study.

In this case, time evolution of the state population under our study is an exponential function
$$[Ar^*](t) = N_0 \exp(-k_d t) \tag{5}$$
with the decay rate of
$$k_d = k_1[Ar][He] + k_2[He]^2 + k_3[He]. \tag{6}$$
Here $k_1$ and $k_2$ are the rate constants for excimerization in reactions (1) and (2), respectively, and $k_3$ is the rate constant for two-body reaction (3).

When monochromatic radiation at the wavelength of 912.3 nm passes the e-beam excited medium, the linear absorption coefficient $k$ should be proportional to the density of the absorbing $4s[3/2]_2^o$ state:
$$k(t) \sim [Ar^*](t). \tag{7}$$
In our experiments, the width of input and output slits of the monochromator was ~0.2 mm providing a satisfactory signal-to-noise ratio, and the width of the monochromator instrumental function was much broader than absorption profile width of the line under study in the gas pressure range of 1.75–4.0 atm. In this case, the Bouguer-Lambert-Beer law does not generally hold, and one has to use an empirical, or so-called modified form of the Bouguer-Lambert-Beer law [18, 19] relating the measured transmission coefficient $T$ to the absorption coefficient $k$ as
$$\ln(1/T) = (kL)\gamma. \tag{8}$$
Here $L$ is the length of the e-beam-excited absorbing medium and $\gamma$ is dimensionless factor which depends on the ratio between the widths of the absorption line and the monochromator instrumental function. Plotting the experimental dependences
$$\ln \ln[1/T(L)] = \text{const} + \gamma \ln L \tag{9}$$

proved applicability of the modified Bouguer-Lambert-Beer law under our conditions and allowed us to measure the dimensionless factor $\gamma$, which happened to be $\gamma = 0.5$ in our experiments (for more details see [20]).

It follows from (8), taking into account relation (7) and the time dependence $[Ar^*](t)$ expected in accordance with (5), that the trailing edge of the "absorption pulse" $\ln(1/T)$ should be strictly exponential:

$$\ln[1/T(t)] \sim \exp(-\gamma k_d\, t). \tag{10}$$

Taking the natural logarithm of both sides of the expression (10) results in the following linear expression for the time evolution of the transmittance $T$ in the e-beam afterglow:

$$\ln\ln[1/T(t)] = \text{const} - \gamma k_d\, t. \tag{11}$$

Typical oscilloscope traces of the probe pulse from the ISI-1 source obtained for He–Ar =100:1 mixture at gas pressure of 2.5 atm for the $4p[1/2]_1 - 4s[3/2]_2^o$ transition wavelength are shown in figure 4. Amplitudes of the signals going into (*B*) and out of (*A*) the excited active medium allows one to find the medium transmittance $T$ at the wavelength selected by the monochromator at each instant of time. Time evolution of the absorption pulse $\ln(1/T)$ in the excited He–Ar mixture derived from the oscilloscope traces in figure 4 is shown in figure 5a. The time evolution of the double logarithm $\ln\ln(1/T)$ at trailing edge of the absorption pulse is shown in figure 5b. One can see that the plotted dependence has a convex-concave shape and cannot be reasonably approximated by linear function (11) (bold line *A* in figure 5b).

Such a time behavior of the $\ln\ln(1/T)$ function has two reasons. Firstly, as in [11, 12], collisional quenching of Ar(4s) level under study is so slow that it is necessary to take into account an influence on its population of even weak recombination flow that continue to populate this level in the afterglow within certain time after termination of the e-beam pulse. In the experiment, such recombination flow manifests itself in appearance of certain convexity within the beginning part of the bold curve *B* in figure 5b.

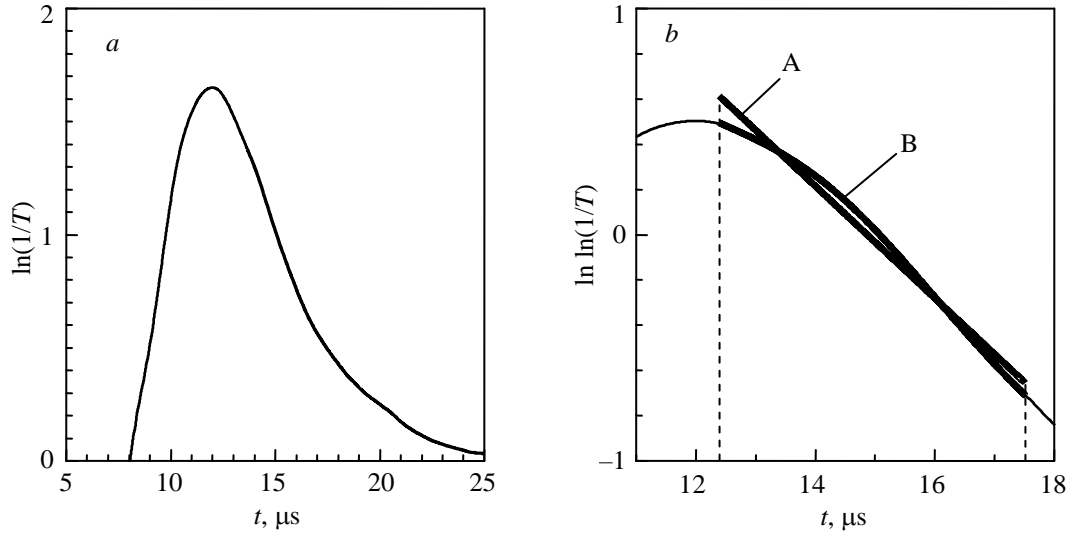

**Figure 5.** Time evolution of the absorption pulse as functions $\ln(1/T)$ (*a*) and $\ln\ln(1/T)$ (*b*) approximated by direct line $\ln(p\exp(-\gamma k_d\,(t-t_0)))$ (bold line *A*) and by the curve

$$\ln(p_{ex}\exp(-k_{ex}\,(t-t_0)) + p_d\exp(-(\gamma g\,(t-t_0))^2)\exp(-\gamma k_d\,(t-t_0)))$$

(bold curve *B*), for the oscilloscope traces in figure 4.

Secondly, the wavelength of $4p[1/2]_1 - 4s[3/2]_2^o$ atomic transition turned out to lie within a broad absorption band of excimer states (see figure 6) whose decay rates are lower than quenching rates of the argon level under study. This leads to additional absorption of the probe pulse and appearance of a slow "tail" (certain concavity) on the bold curve *B* in figure 5b (see also [13]).

To "extract" the sought-for exponential component in the absorption pulse trailing edge with $k_d$ set by the reactions (1)–(3), it was proposed in [13] to numerically simulate the time dependences like that presented in figure 5b using the following approximation:

$$\ln\ln(1/T) = \ln(p_{ex}\exp(-k_{ex}\,(t-t_0)) + p_d\exp(-(\gamma g\,(t-t_0))^2)\exp(-\gamma k_d\,(t-t_0))), \tag{12}$$

where $(t-t_0)$ is time counted from the moment of e-beam pulse termination $t_0$; $p_{ex}$, $p_d$, $k_{ex}$, $g$ and $k_d$ are the approximation parameters.

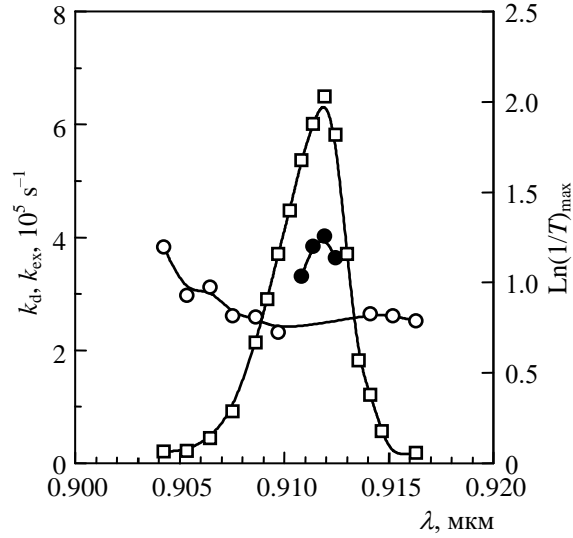

**Figure 6.** Decay rates $k_d$ for argon $4s[3/2]_2^o$ state (●) and $k_{ex}$ for excimer state (○), and the absorption pulse amplitudes $\ln(1/T)_{max}$ (□) obtained from the data of absorption measurements in the vicinity of $4p[1/2]_1 - 4s[3/2]_2^o$ argon transition in He–Ar=100:1 mixture at gas pressure of 3.0 atm.

To find the decay rate $k_d$ from the measured $\ln\ln(1/T)$ time dependence, it was simulated by using the procedure of least squares by varying $p_{ex}$, $p_d$, $k_{ex}$, $g$ and $k_d$ parameters in accordance with the Levenberg–Marquardt algorithm (LMA) [21, 22]). A set of experimental data for the mixtures He–Ar = 50:1, 75:1, 100:1 and 200:1 at gas pressures from p = 1.75 to 4.0 atm was numerically calculated with a step of 0.25 atm.
In the coordinates of figure 5a the function (12) corresponds to the function
$$\ln(1/T) = p_{ex} \exp(-k_{ex}(t-t_0)) + p_d \exp(-(\gamma g(t-t_0))^2) \exp(-\gamma k_d(t-t_0)), \tag{13}$$
where the first, slowly-decaying term of the right-hand side of (13)
$$p_{ex} \exp(-k_{ex}(t-t_0)) \tag{14}$$
describes a broadband absorption ($\gamma = 1$) of the probe pulse by the excimer molecules, whereas the second term
$$p_d \exp(-(\gamma g(t-t_0))^2) \exp(-\gamma k_d(t-t_0)) \tag{15}$$
describes the studied exponential process of collisional quenching of $4s[3/2]_2^o$ metastable state
$$\exp(-\gamma k_d(t-t_0)) \tag{16}$$
with "Gaussian pre-exponent"
$$\exp(-(\gamma g(t-t_0))^2), \tag{17}$$
which provides a correction for the recombination and relaxation processes of populating this state [11, 12]. Obtaining a true form of the Gaussian pre-exponent in analytical form is hardly possible because of the variety and complexity of the relevant reactions depending, in particular, on the time-varying temperature and density of the secondary electrons. However, in practice, the above description gives satisfactory results within the transmittance dynamic range $T = 0.1 - 0.9$ (see figure 4). To illustrate, in figure 7 for each He–Ar mixture the values of $k_d$ [He]$^{-1}$, i.e., the quenching rate reduced to He buffer-gas concentration, are plotted as a function of this concentration. In accordance with linear dependence expected from (6)
$$k_d[\text{He}]^{-1} = (\delta k_1 + k_2)[\text{He}] + k_3 \tag{18}$$
(here $\delta = [\text{Ar}]/[\text{He}]$) the experimental points with good accuracy lie on the straight lines originating from a common point on the ordinate axis, which proves adequacy of the representation (12) and (13) and correctness of the described procedure of experimental data processing.

The obtained set of experimental $k_d^i$ ([Ar], [He]) points was used to evaluate the rate constants for plasma-chemical reactions (1)–(3) (see Table 1). The rate constants $k_1$, $k_2$, $k_3$ presented in Table 1 were evaluated using the method of least squares and the Levenberg–Marquardt algorithm by varying the sought-for rate constants in
$$k_d^i = k_1[\text{Ar}][\text{He}] + k_2[\text{He}]^2 + k_3[\text{He}] \tag{19}$$

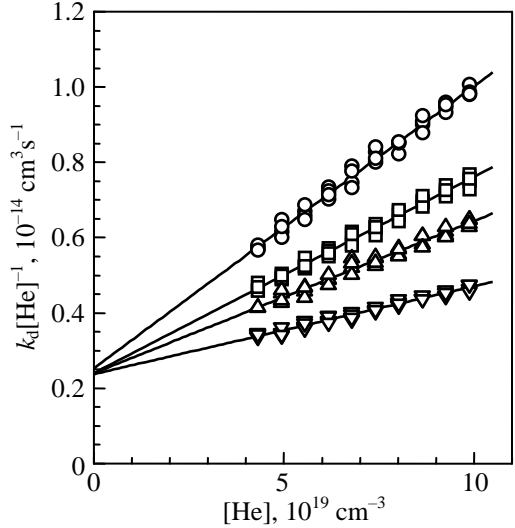

**Figure 7.** Reduced quenching rate $k_d [\text{He}]^{-1}$ of $4s[3/2]_2^o$ argon state versus helium concentration for mixtures He–Ar = 50:1 (○), 75:1 (□), 100:1 (△) and 200:1 (▽).

simultaneously for the whole set of experimental data $k_d^i$. The core of the procedure consists in construction in the coordinates ([He], [Ar]) of the curvilinear surface

$$S([\text{He}], [\text{Ar}]) = k_1[\text{He}][\text{Ar}] + k_2[\text{Ar}]^2 + k_3[\text{Ar}], \tag{20}$$

having the smallest deviation, in terms of the procedure, from the set of the experimental points (see [13] for details).

**Table. 1.** Measured rate constants for collisional quenching of the $4s[3/2]_2^o$ argon state in He–Ar-mixtures.

| Reaction | Upper bounds | Rate constant |
|---|---|---|
| (1) | $3.8 \times 10^{-33}$ cm$^6$/s | $(3.6 \pm 0.4) \times 10^{-33}$ cm$^6$/s |
| (2) | $< 10^{-35}$ cm$^6$/s | $(4.4 \pm 0.9) \times 10^{-36}$ cm$^6$/s |
| (3) | $7.2 \times 10^{-15}$ cm$^3$/s | $(2.4 \pm 0.3) \times 10^{-15}$ cm$^3$/s |

Processing of the experimental dependences $\ln\ln[1/T(t)]$ within the above-described technique but using a simpler linear approximation (the bold line *A* in figure 5b, see also [14]) allowed us to estimate the reaction rate constants with rather limited accuracy, which estimate makes it possible to find an "upper bound" for the rate constants (Table 1).

The results obtained in the present study demonstrate that under our experimental conditions collisional quenching of 4s states of the Ar atom in He–Ar mixture occurs through the three-body reaction with formation of $\text{Ar}_2^*$ homonuclear dimer (1) and two-body reaction (3). The three-body reaction (2) with formation of HeAr$^*$ heteronuclear dimer hardly contribute to collisional quenching of Ar*. This may be related to the fact that heteronuclear dimer HeAr$^*$, a product of reaction (2), turns out to be unstable because of its low binding energy [23] and rapidly decomposes into the initial components in the reverse collisional reaction. Thus, the measured effective rate constant for the formation of HeAr* dimer happens to be negligible.

## 3. Conclusion

A new approach to the development of high-power high-pressure quasi-cw He–Ar laser with a large volume of gain medium at $4p[1/2]_1 - 4s[3/2]_2^o$ ArI transition with electron beam-optical pumping is proposed in this paper. It is proposed to create and maintain the necessary density of excited metastable states $4s[3/2]_2^o$ in the gain medium using a fast electron beam and then to pump the upper laser level by a laser diode in accordance with the classical three-level scheme. The offered technique does not set a fundamental limitation on the excited volume dimensions and lasing pulse width.

Collisional quenching of lower laser level $4s[3/2]_2^o$ in He–Ar mixtures excited by a beam of fast electrons was experimentally studied. The rate constants for plasma-chemical reactions (1)–(3) are measured for the first time by the procedure of absorption probing. It is shown that the main channels of quenching $4s[3/2]_2^o$ state are the processes of excimerization with formation of $Ar_2^*$ dimer with rate constant $3.6 \times 10^{-33}$ cm$^6$/s and quenching by buffer gas with rate constant $2.4 \times 10^{-15}$ cm$^3$/s, whereas the reaction with formation of the heteronuclear dimer plays virtually no role. The measurements demonstrated that the rate of collisional quenching of $4s[3/2]_2^o$ state is low being considerably inferior to the rates of similar processes in the mixtures of other rare gases measured earlier in [10–16]. This circumstance does indicate that production and maintenance of the necessary amount of argon atoms in $4s[3/2]_2^o$ state is feasible at very moderate current densities of pumping electron beam readily attainable with, e.g., hot-cathode electron guns whose pulse width can be easily regulated up to a cw mode of operation. Experimentally demonstrated long lifetime of metastable argon up to tens of microseconds at pressures of e-beam-excited He–Ar mixtures up to several atmospheres can lead, under electron beam-optical pumping, to a comparably long-lasting population inversion and laser pulse. Employment of appropriate cw e-beam and optical pumping sources can make possible creation of a cw large-volume high-pressure near-IR He-Ar laser.

This research was supported by the Russian Foundation for Basic Research (Project 17-02-00241).

## 4. References

[1]  Chapovsky P L, Lisitsyn V N and Sorokin A R 1976 *Opt. Comm.* **16** (1) 33
[2]  Han J and Heaven M 2012 *Opt. Lett*. **37** 2157
[3]  Han J, Glebov L, Venus G and Heaven 2013 *Opt. Lett*. **38** 5458
[4]  Han J and Heaven M 2014 *Opt. Lett*. **39** 6541
[5]  Basov N G, Chugunov A Y, Danilychev V A, Kholin I V and Ustinovsky N N 1983 *IEEE J. Quant. Electron.* **QE-19** (2) 126
[6]  Basov N G, Baranov V V, Chugunov A Y, Danilychev V A, Dudin A Y., Kholin I V, Ustinovsky N N and Zayarnyi D A 1985 *IEEE J. Quant. Electron.* **QE-21** (11) 1756
[7]  N.G. Basov, Baranov V V, Danilychev V A, Dudin A Y., Zayarnyi D A, Rzhevskiĭ A V, Ustinovsky N N, Kholin I V and Chugunov A Y 1986 *Soviet J. of Quant. Electron.* **16** (8) 1008
[8]  Kholin I V 2003 *Quant. Electron.* **33** (2) 129
[9]  Zayarnyi D A and Kholin I V 2003 *Quantum Electronics* **33** (6) 474
[10] L.V. Semenova, Ustinovsky N N and Kholin I V 2004 *Quant. Electron.* **34** (3) 189
[11] Zayarnyi D A, L'dov A Yu and Kholin I V 2009 *Quant. Electron.* **39** (9) 821
[12] Zayarnyi D A, L'dov A Yu and Kholin I V 2010 *Quant. Electron.* **40** (2) 144
[13] Zayarnyi D A, L'dov A Yu and Kholin I V 2011 *Quant. Electron.* **41** (2) 128
[14] Zayarnyi D A, L'dov A Yu and Kholin I V 2013 *Quant. Electron.* **43** (8) 720
[15] Zayarnyi D A, L'dov A Yu and Kholin I V 2014 *Quant. Electron.* **44** (11) 1066
[16] Kholin I V, L'dov A Yu, Ustinovsky N N and Zayarnyi D A 2015 *J. of Russian Laser Research* **36** (1) 74
[17] Zayarnyi D A, L'dov A Yu, Ustinovsky N N and Kholin I V 2010 *Instr.and Exp. Tech.* **53** (4) 561
[18] Oka T 1977 *Res. Rep. Nagaoka Tech. Coll.* **13** (4) 207
[19] Davis C C and McFarlane R A 1977 *J. Quant. Spectrosc. Radiat. Transfer* **18** 151
[20] Zayarnyi D A, Kholin I V and Chugunov A Y 1995 *Quantum Electronics* **25** (3) 217
[21] Levenberg K 1944 *Quarterly of Appl. Math.* **2** 164
[22] Marquardt D 1963 *SIAM J. on Appl. Math.*, **11** (2), 431
[23] Nowak G and Fricke J 1985 *J. Phys. B: At. Mol. Phys*. **18** (7) 1355